\documentclass[conference]{IEEEtran}
\usepackage{amsmath,amssymb,amsfonts}
\usepackage{graphicx}
\usepackage{float}
\usepackage{subfig}
\usepackage{color}
\usepackage{cite}
\usepackage{multirow}
\usepackage{algorithm}  
\usepackage{algorithmicx}  
\usepackage{algpseudocode} 
\usepackage{bm}
\usepackage[colorlinks,linkcolor=black,anchorcolor=black,
citecolor=black,urlcolor=black]{hyperref}

\usepackage{stfloats}
\usepackage{siunitx}
\DeclareSIUnit\bps{bps}
\DeclareSIUnit\Torr{Torr}
\DeclareSIUnit\torr{Torr}
\DeclareSIUnit\sample{Sa}
\newcommand*{\circled}[1]{\lower.7ex\hbox{\tikz\draw (0pt, 0pt)%
  circle (.5em) node {\makebox[1em][c]{\small #1}};}}

\ifCLASSINFOpdf

\else

\fi

\graphicspath{{figures/}}

\begin{document}

\title{Correlation-based Dual-band THz Channel Measurements and Characterization in a Laboratory}

\author{
\IEEEauthorblockN{Yuanbo~Li\IEEEauthorrefmark{1}, Yiqin Wang\IEEEauthorrefmark{1}, Yi Chen\IEEEauthorrefmark{2}, Ziming Yu\IEEEauthorrefmark{2}, and Chong~Han\IEEEauthorrefmark{1}}
\IEEEauthorblockA{\IEEEauthorrefmark{1} Terahertz Wireless Communications (TWC) Laboratory, Shanghai Jiao Tong University, China \\ Email:  \{yuanbo.li,wangyiqin,chong.han\}@sjtu.edu.cn\\
\IEEEauthorrefmark{2} Huawei Technologies Co., Ltd, China.
Email: \{chenyi171,yuziming\}@huawei.com
}
}

\maketitle

\begin{abstract}
The Terahertz band, spanning from 0.1~THz to 10~THz, is envisioned as a key technology to realize ultra-high data rates in the 6G and beyond mobile communication systems, due to its abundant bandwidth resource. However, to realize THz communications, one substantial step is to fully understand the THz channels, which relies on extensive channel measurements. In this paper, using a correlation-based time domain channel sounder, measurement campaigns are conducted in a laboratory at 140~GHz and 220~GHz. In the data post-processing procedures, the time drift of clock signals is corrected using a linear interpolation/extrapolation method. Based on the measured results, the main objects that provide significant once-scattering clusters are found, based on which the scattering losses are calculated and analyzed. Furthermore, the channel characteristics, including path loss, shadow fading, K-factor, etc. are calculated and compared to 3GPP standard values. The propagation analysis and channel characteristics are helpful to study channel modeling and guide system design for THz communications.
\end{abstract}

\IEEEpeerreviewmaketitle

\section{Introduction}
\par The sixth generation (6G) and beyond mobile communication network is visioned to enable plenty of thrilling applications, such as metaverse, digital twin, among others~\cite{chen2021terahertz}. In light of this, ultra-high data rates, e.g., 1 Terabits per second, are required to support the explosively grown data traffic among intelligent devices and applications. To realize ultra-high-speed communications, the Terahertz (THz) band, ranging from \SI{0.1}{THz} to \SI{10}{THz}, is envisioned as a key technology owning to its abundant yet unregulated spectrum resource and ultra-large contiguous usable bandwidth~\cite{Akyildiz2022Terahertz}. 
\par However, to realize THz communications, one major challenge lies on channel modeling in the THz band, which requires extensive measurement efforts to analyze new propagation phenomena and extract new channel characteristics. With maturity of THz hardware, recently a few research groups have built up THz channel sounders and conducted channel measurement campaigns. For example, the research group from Technische Universität Braunschweig has reported many measurement results since 2011, investigating conference room, office, as well as vehicle-to-vehicle channels~\cite{priebe2011channel,priebe2013ultra,Eckhardt2021channel}. In addition, the research group at New York University has conducted massive channel measurements in \SIrange{140}{142}{GHz} band in various scenarios, including office, factory, urban, etc.~\cite{ju2021millimeter,ju2022sub,ju2021sub}. Moreover, the research group from University of Southern California has made progress in channel characterization in frequency bands near \SI{140}{GHz} and \SI{220}{GHz} in outdoor urban scenarios~\cite{abbasi2020double,abbasi2021double,Abbasi2023THz}. Based on a vector network analyzer (VNA)-based method, we have conducted measurements in both indoor and outdoor scenarios at Shanghai Jiao Tong University, from \SI{140}{GHz} up to \SI{400}{GHz}~\cite{chen2021channel,wang2022thz,li2022channel,li2023channel,wang2023thz}. 
\par In this paper, by using a correlation-based time-domain channel sounder, which can achieve long measurable distance ($>$\SI{200}{m}) and ms-level measurement duration for one channel impulse response (CIR), dual-band channel measurement campaigns are conducted in a laboratory at \SI{140}{GHz} and \SI{220}{GHz}. There are various objects in the laboratory, such as experimental tables, racks, desks, etc., for which it is measured to investigate the THz wave propagation under complex environments, which were not analyzed in the aforementioned literature. Overall, 10 Rx positions are measured with 3600 CIRs to characterize the THz wave propagation in the laboratory. To obtain accurate delay of MPCs, the time drift of the clocks is corrected based on a linear interpolation/extrapolation method. 

To analyze the propagation features in the laboratory in the dual THz bands, the influences of objects are carefully analyzed, where the scattering loss is calculated and analyzed. Furthermore, channel characteristics, including the path loss, shadow fading, K-factor, etc., are investigated and compared with existing 3GPP standardized channel model. The measured results not only reveal the strong sparsity and weak multipath effects of THz channels, but also establish numerology that is helpful for THz communication system designs.
\par The remainder of the paper is organized as follows. In Sec.~\ref{sec:measurement}, the correlation-based channel sounder, measurement set-up and deployment are explained in detail. Furthermore, the data processing procedure is introduced in Sec.~\ref{sec:timedrift}. In light of the measurement results, the propagation analysis and channel characteristics are elaborated in Sec.~\ref{sec:char}. Finally, Sec.~\ref{sec:conclude} concludes the paper. 

\section{Channel Measurement Campaign}
\label{sec:measurement}
\par  In this section, the measurement system, set-up and deployment in the laboratory is described.
\subsection{Measurement System}
\label{sec:system}
\begin{figure}
    \centering
    \subfloat[Diagram of the channel sounder.] {
     \label{fig:sounder}     
    \includegraphics[width=0.9\columnwidth]{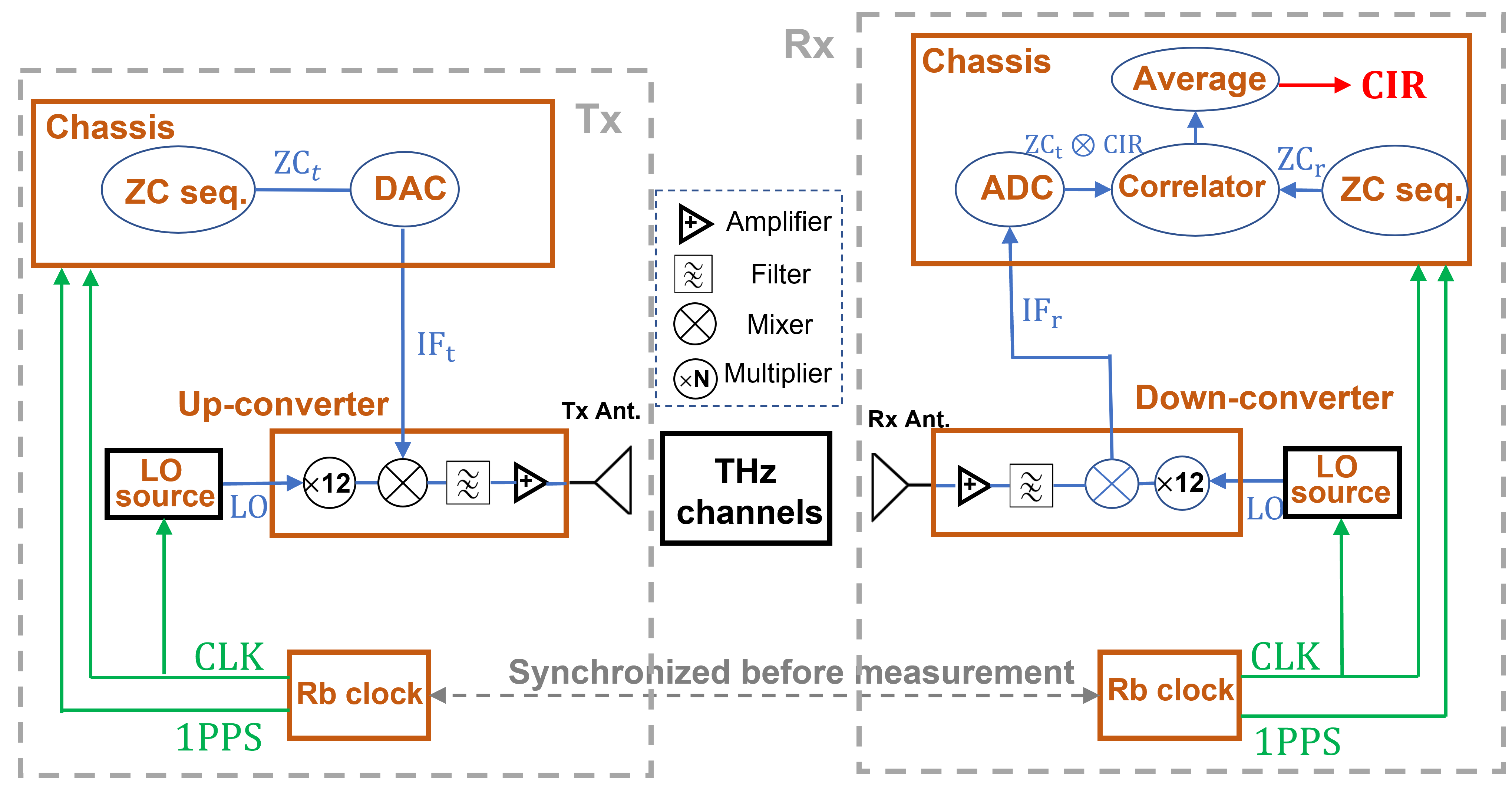}  
    }
    \quad
    \subfloat[Picture of hardware.] {
     \label{fig:hardware1}     
    \includegraphics[width=0.9\columnwidth]{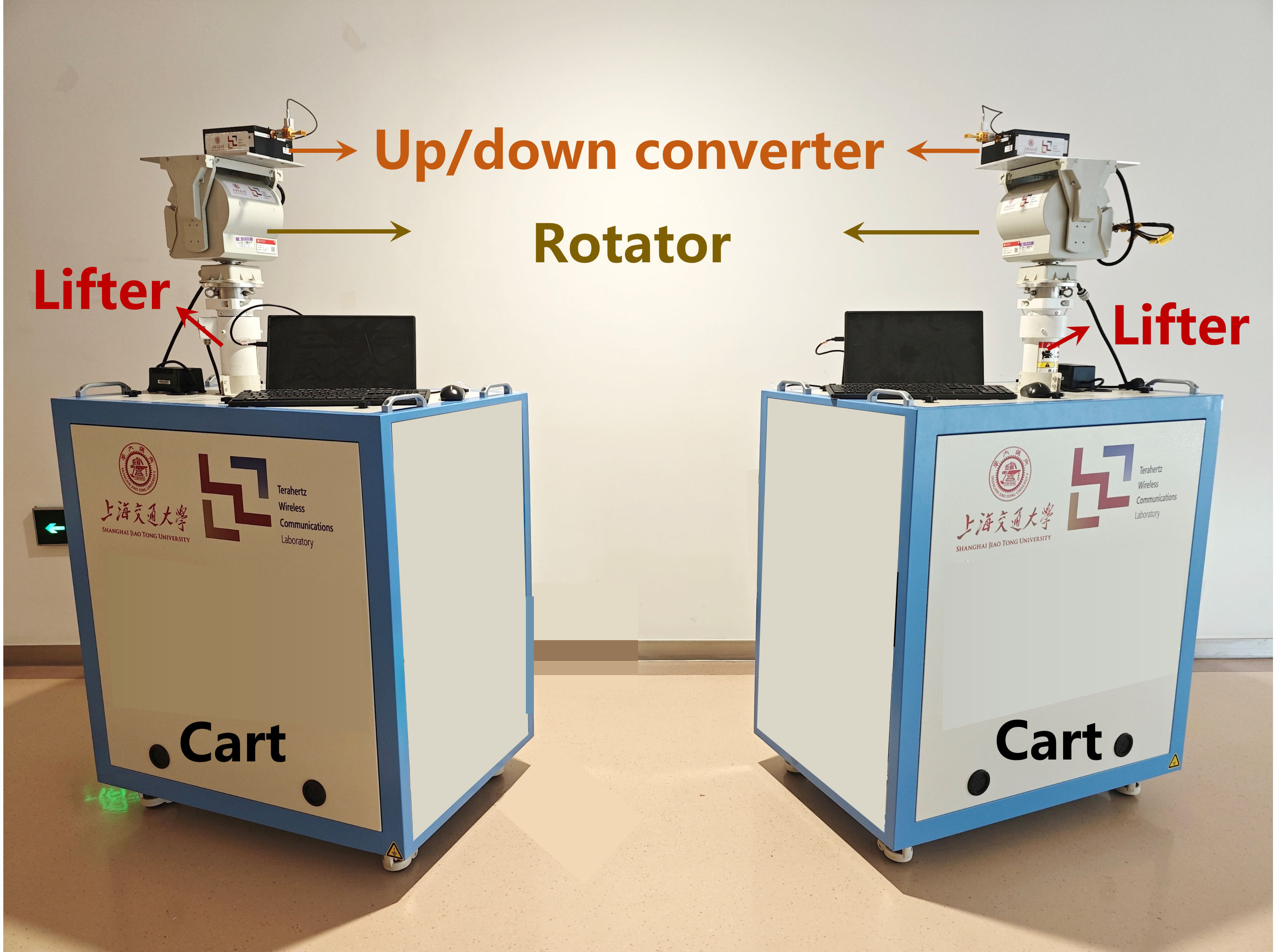}  
    }
    \caption{Correlation-based time-domain channel sounder.}
    \label{fig:cs}
    \vspace{-0.5cm}
\end{figure}
\par The diagram of the channel sounder is shwon in Fig.~\ref{fig:cs}(a). The channel sounder consists of a transmitter (Tx) and a receiver (Rx). Both the Tx and Rx include a chassis and a radio frequency (RF) front end. A Zadoff-Chu (ZC) sequence is firstly generated in the Tx chassis. The ZC sequence passes through a digital-to-analog converter (DAC) and up-converted to a intermediate frequency (IF) at \SI{12}{GHz}. The IF signal is then sent to the Tx RF front end, namely the up-converter, and mixed with a local oscillation (LO) signal to produce the THz signal, which is further band-pass filtered and amplified. The THz signal radiates out through the Tx antenna, propagates through the THz channel, and is then received at the Rx side. Through a reverse procedure as that in the Tx, i.e., amplifying, filtering, down-converting and analog-to-digital conversion, the digital bits can be decoded, which contain the convolution of the ZC sequence and the channel impulse response (CIR). By correlating the digital bits with an exactly same ZC sequence, one CIR sample can be obtained. Finally, many samples are averaged to increase the signal-to-noise (SNR) ratio, after which the CIR is obtained and recorded.
\par Apart from the aforementioned equipment, rubidium (Rb) clocks are used to provide reference frequency for the LO sources, as well as clock and 1 pulse-per-second (1PPS) trigger signal to the chassis. Before conducting channel measurements, the Rb clocks at Tx and Rx are connected in a master-slave mode to synchronize the 1PPS signal, which may last several hours. After the clock synchronization, the Tx and Rx are placed in interested positions to measure the THz channels. During the measurement process, the synchronized clock signal would linearly drift as the time goes by, e.g., the time difference between the two 1PPS signal may increase by \SIrange{10}{30}{ns} per hour, for which a linear interpolation/extrapolation method is utilized to correct the time drift in the data post-processing procedure, as explained in~Sec.\ref{sec:timedrift}. Furthermore, as shown in Fig.~\ref{fig:cs}(b), the up/down converters are installed on rotators, lifters, and carts, to conveniently change their steering directions, heights, and positions, respectively. 
\par Different from the VNA-based channel sounder~\cite{chen2021channel,wang2022thz,li2022channel,li2023channel,wang2023thz}, no cable connection is needed during the channel measurements using the correlation-based sounder. Through a 1000-times average, the noise floor is lower than \SI{160}{dB}, for which the measurable distance with a \SI{30}{dB} SNR in the line-of-sight (LoS) case exceeds \SI{200}{m}. Moreover, the measurement duration for recording one CIR is around \SI{6}{ms}.
\subsection{Measurement Setup}

  

  
    


            
            
        
		




\par The measurement setups are introduced in detail as follows. Two frequency bands centered at \SI{140}{GHz} and \SI{220}{GHz} are measured, with a bandwidth of \SI{1.536}{GHz} band. Correspondingly, the time resolution is \SI{0.65}{ns}, which means that any two MPCs with propagation distance difference larger than \SI{19.5}{cm} can be distinguished. Moreover, the measured CIRs contain 2048 sampling points, resulting in a \SI{1332.7}{ns} maximum measurable delay and a \SI{399.8}{m} maximum path length. The heights of Tx and Rx are set as \SI{2.5}{m} and \SI{1.6}{m}, respectively. 
\par Moreover, the transmitter directly radiates out THz waves through a standard waveguide WR5, which has \SI{7}{dBi} antenna gain and a $30^\circ$ half-power beamwidth (HPBW). By contrast, the Rx is equipped with a horn antenna with a \SI{25}{dBi} gain and a $8^\circ$ HPBW. To capture MPCs from various directions, direction-scan sounding (DSS) is conducted by mechanically rotating the Rx to scan the azimuth plane from $0^\circ$ to $360^\circ$ and elevation plane $-20^\circ$ to $20^\circ$, with a $10^\circ$ angle step. {Therefore, there are totally 180 CIRs measured at each Rx position.}
\subsection{Measurement Deployment}
\begin{figure}
    \centering
    \includegraphics[width=1.0\columnwidth]{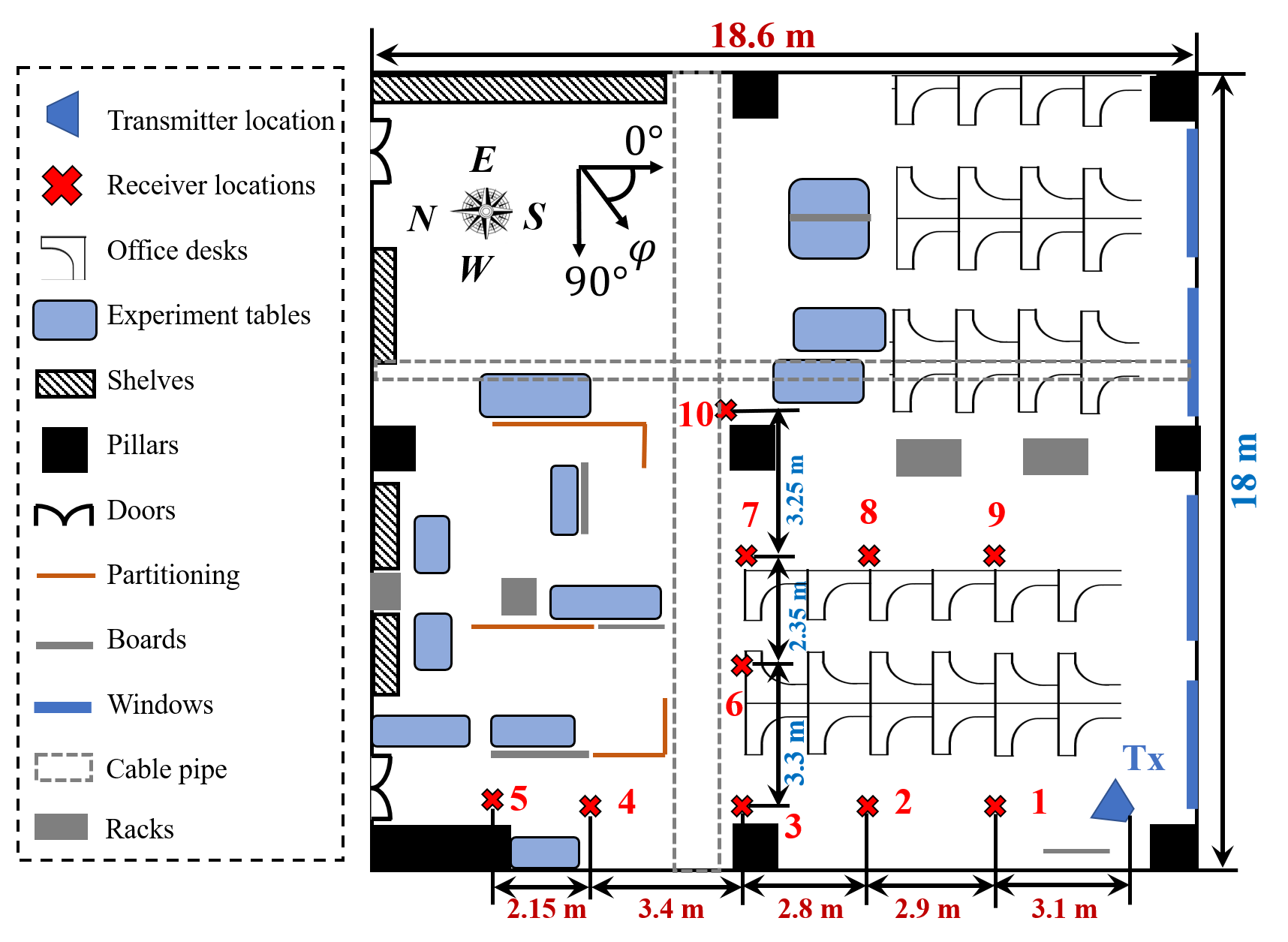}  
    \caption{The measurement deployment in the laboratory.}
    \label{fig:layout}
    \vspace{-0.5cm}
\end{figure}
\par The measurement campaign is conducted in a laboratory in the Longbin Building at Shanghai Jiao Tong University, as shown in Fig.~\ref{fig:layout}. The laboratory can be roughly divided into an office area (the southern half) and an experimental area (the northern half). In the experimental area, there are plenty of experimental tables, shelves, racks, etc., while main objects in the office area are office desks. Moreover, the ceiling of the laboratory is furnished with metal pipes to route the electric power. Furthermore, The length and width of the laboratory are \SI{18.6}{m} and \SI{18}{m}, respectively.
\par The transmitter remains static, deployed near southwest corner in the laboratory. Besides, 10 receiver positions are measured, whose positions are shown in Fig.~\ref{fig:layout}. When measuring each Rx position, the steering direction of the Tx is adjusted to directly point at the receiver. The separation distance between Tx and Rx ranges from \SI{3}{m} to \SI{14}{m}. Moreover, only point 10 is in the none-line-of-sight (NLoS) region, while other Rx locations have LoS propagation. {For each Rx point, it takes around 20 minutes to measure the channel. Overall, there are 3600 CIRs that are measured.}
\section{Data Post-processing Procedures and Time Drift Correction}
\label{sec:timedrift}
The data post-processing procedures include calibration, time drift correction, channel estimation, and MPC clustering. To begin with, since the measured CIRs include both the channel responses and the system responses, a direct-connection measurement is conducted to measure only the system responses, which is further eliminated from the real measurement data through the calibration process. Moreover, as the timing difference of the two 1PPS signals of the two Rb clocks would gradually increase, a time drift correction process is then implemented obtain CIRs with accurate absolute delay. What's more, the MPC parameters are estimated, based on which the main clusters are found through MPC clustering. 
\par While the calibration, channel estimation, and MPC clustering procedures are the same as those in our previous work~\cite{li2023channel}, one unique challenge for the correlation-based channel sounder is the correction of time drift of the Rb clocks.
\label{sec:dataprocess}
In our measurements, the two Rb clocks providing clock and trigger signals to Tx and Rx are synchronized in a master-slave mode for 1.5~hour, after which they are disconnected and the measurement campaigns are conducted. During the measurement process, the 1PPS signal of the two Rb clocks will gradually drift away from each other, resulting in inaccuracy of the time delay measured in CIRs, which is termed as \textit{time drift}~\cite{MacCartney2017Flexible}. To obtain the absolute delay of MPCs, the time drift must be corrected.
\begin{figure}
    \centering
    \includegraphics[width=0.8\columnwidth]{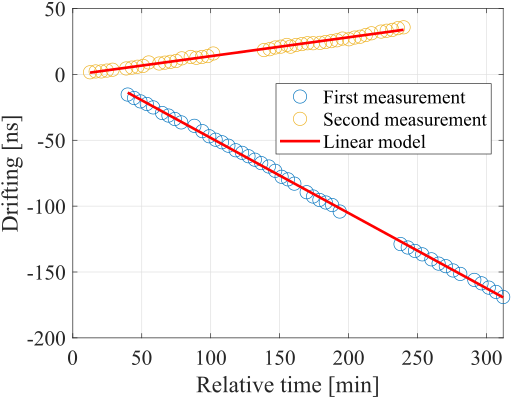}
    \caption{The time drift observed during measurement. The relative time starts from the time instant when the direct-connection measurement is conducted.}
    \label{fig:rb}
    \vspace{-0.5cm}
\end{figure}
\par Since the Rx points 1-9 are all in LoS area, the theoretical delay of the LoS path can be calculated based on the positions of Tx and Rx, while the measured delay of the LoS path can be easily obtained as the delay of the strongest path when the Rx is in the reference scanning direction, i.e., the steering direction towards the Tx. Therefore, the time drift is evaluated as the difference between the theoretical delay and measured delay of the LoS path, which is shown in Fig.~\ref{fig:rb}. Two independent measurement campaigns are conducted to measure the two frequency bands \SI{140}{GHz} and \SI{220}{GHz}. As can be observed, in both measurements, the time drift grows linearly as the relative time increases. Therefore, the time drift can be calculated and corrected based on linear interpolation/extrapolation, as
\begin{equation}
    \Delta\tau(t)=
    \begin{cases}
        \frac{(t-t_i)(\Delta\tau_j-\Delta\tau_i)}{t_j-t_i}+\Delta\tau_j,&t_i<t<t_j, i,j\in[1,L]\\
        at+b,&t<t_1~\text{or}~t>t_L
    \end{cases}
\end{equation}
where $\Delta\tau_1,\Delta\tau_2,...,\Delta\tau_L$ are the time drift observed in the reference direction at relative time $t_1,t_2,...,t_L$, respectively, with $L$ denoting the overall number of CIRs measured in the reference direction. For linear interpolation when the time $t$ is within the range $[t_1,t_L]$, $t_i$ and $t_j$ are two time instants measured in the reference direction that are closest to $t$. Furthermore, when $t$ is outside of the range $[t_1,t_L]$, a linear model is used for extrapolation, where the parameters $a$ and $b$ are obtained by conducting linear regression using the observed time drift in reference directions.
\section{Propagation Analysis and Channel Characterization}
\label{sec:char}
\par In this section, the propagation analysis and channel characteristics are elaborated. To begin with, a thorough propagation analysis is conducted, identifying the objects in the laboratory that significantly affect the THz wave propagation. Moreover, the channel characteristics are calculated and analyzed. Since only one Rx point is measured in the NLoS region, the channel characteristics in the LoS case are focused and studied.
\subsection{Propagation Analysis}
\label{sec:prop}
\begin{figure*}
    \centering
    \includegraphics[width=1.6\columnwidth]{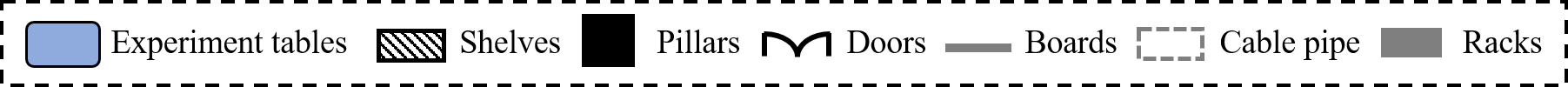}
    \vspace{-0.5cm}
\end{figure*}
\begin{figure*}[!tbp]
    \centering
    \subfloat[\SI{140}{GHz}]{ 
    \includegraphics[width=0.8\columnwidth]{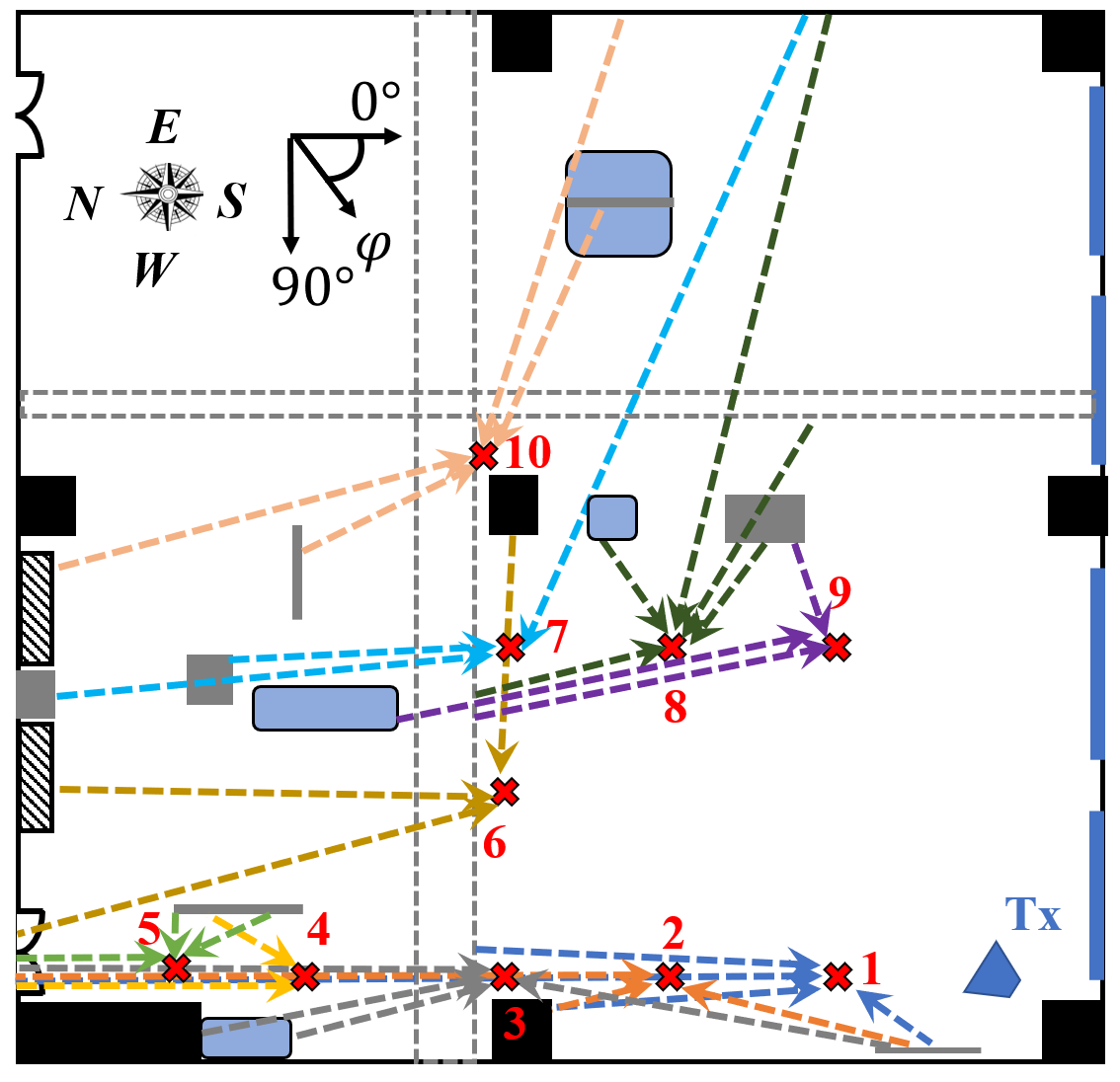}
    }
    \subfloat[\SI{220}{GHz}]{ 
    \includegraphics[width=0.8\columnwidth]{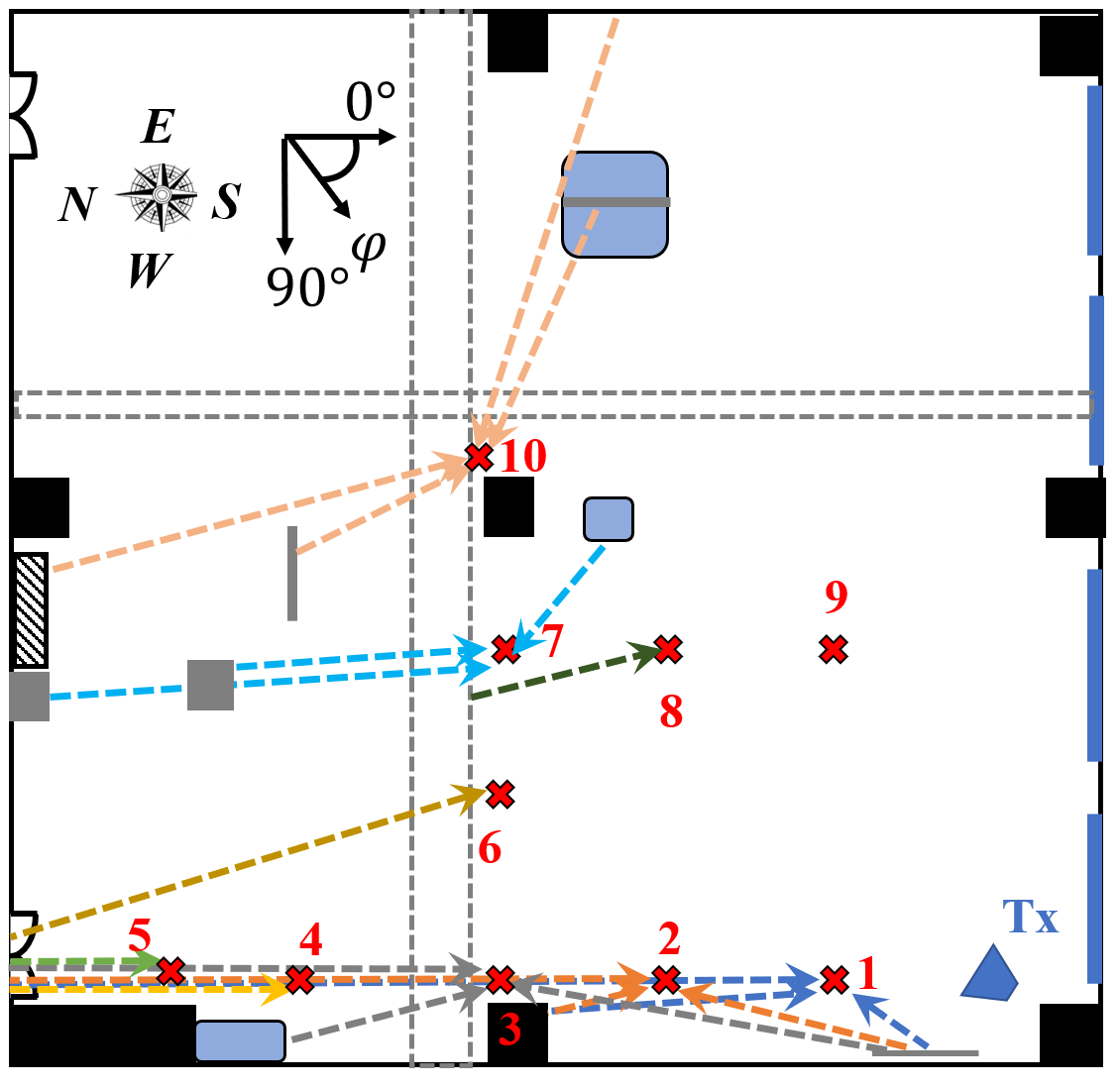}  
    }   
    \caption{Propagation paths of once-scattering clusters. Only the objects that provide once-scattering clusters are shown. For clear illustration, the propagation paths from the Tx to the scatterers are omitted.}
    \label{fig:prop}
    \vspace{-0.5cm}
\end{figure*}
\begin{figure*}[!tbp]
    \centering
    \subfloat[\SI{140}{GHz}]{ 
    \includegraphics[width=0.85\columnwidth]{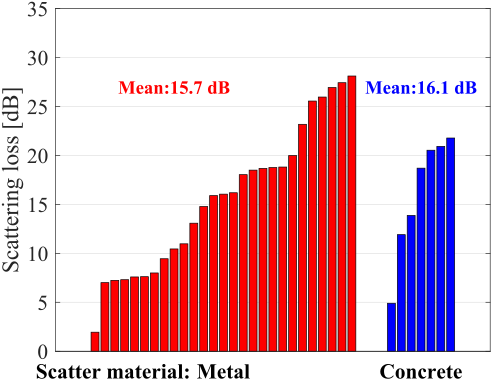}
    }
    \subfloat[\SI{220}{GHz}]{ 
    \includegraphics[width=0.85\columnwidth]{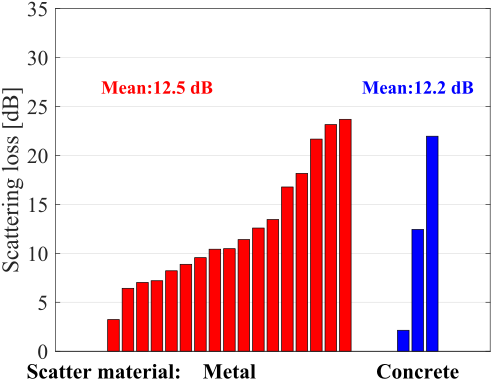}  
    }   
    \caption{Scattering loss of once-scattering clusters. Each bar stands for one cluster.}
    \label{fig:scatloss}
    \vspace{-0.5cm}
\end{figure*}
\par To examine how the objects in the laboratory affect the propagation of THz waves, the propagation paths of the once-scattering clusters are traced based on their delays and direction-of-arrival (DoA), which are shown in Fig.~\ref{fig:prop}. Several observations are made as follows. First, the main objects affecting the THz channels include boards, shelves, racks, experimental tables, walls, pillars, cable pipes, as well as doors, which are mostly made of metal besides the concrete pillars and walls. Second, comparing the two frequency bands, it can be seen that there exists more scatterers providing once scattering in the \SI{140}{GHz} band, compared to the \SI{220}{GHz} band. In fact, some scattering clusters observed in the \SI{140}{GHz} band, such as those observed at Rx point 9, are too weak to be extracted at \SI{220}{GHz}, maybe due to the increase of frequency. Nonetheless, there are also some clusters that exist at \SI{220}{GHz} but vanish at \SI{140}{GHz}, such as that from the experimental table at Rx point 7. Generally speaking, the multipath richness is higher at \SI{140}{GHz} compared to \SI{220}{GHz}.
\par By extracting the propagation delay and path gain of the once-scattering clusters, the scattering loss is calculated as
\begin{equation}
    \text{L}_{\text{Sca}}=-20\log_{10}{\alpha_{\text{Sca}}}-\text{FSPL}(c\tau_{\text{Sca}},f),
\end{equation}
where $\alpha_{\text{Sca}}$ and $\tau_{\text{Sca}}$ are path gain and delay of the once-scattering clusters. Moreover, $\text{FSPL}$ is the free space path loss, which is calculated by using the Friis' law
\begin{equation}
    \text{FSPL}(d,f)=-20\times\log_{10}\frac{c}{4\pi fd},
\end{equation}
where $d$ and $f$ are propagation distance and carrier frequency. Moreover, $c$ denotes the speed of light.
\par The scattering losses of the once-scattering clusters are shown in Fig.~\ref{fig:scatloss}, from which several observations are made as follows. First, most of the once-scattering clusters are originated from metal objects, indicating that the metal objects in indoor environments are main factors affecting the THz wave propagation. Second, the mean scattering loss from metal objects is close to that from concrete objects. The scattering losses from concrete objects are concentrated within the range \SIrange{10}{25}{dB}, while the scattering losses from metal objects are roughly uniformly distributed between \SIrange{2}{25}{dB}. In other words, the metal objects provide not only many significant scattering clusters with scattering loss less than \SI{10}{dB}, but also weak clusters with scattering loss larger than \SI{20}{dB}. Third, comparing the two frequency bands, it can be observed that the mean scattering loss is even smaller at \SI{220}{GHz} than that at \SI{140}{GHz}. However, we clarify that these numbers are not sufficient to lead to strong statements on the frequency dependency of scattering loss, since the scattering loss is also dependent on many other factors, such as material, surface roughness, incident angle, etc.
\subsection{Path Loss and Shadow Fading}
\begin{table}[tbp]
    \vspace{0.1cm}
    \centering
    \caption{Channel characteristic distribution parameters.}
    \includegraphics[width = 0.9\columnwidth]{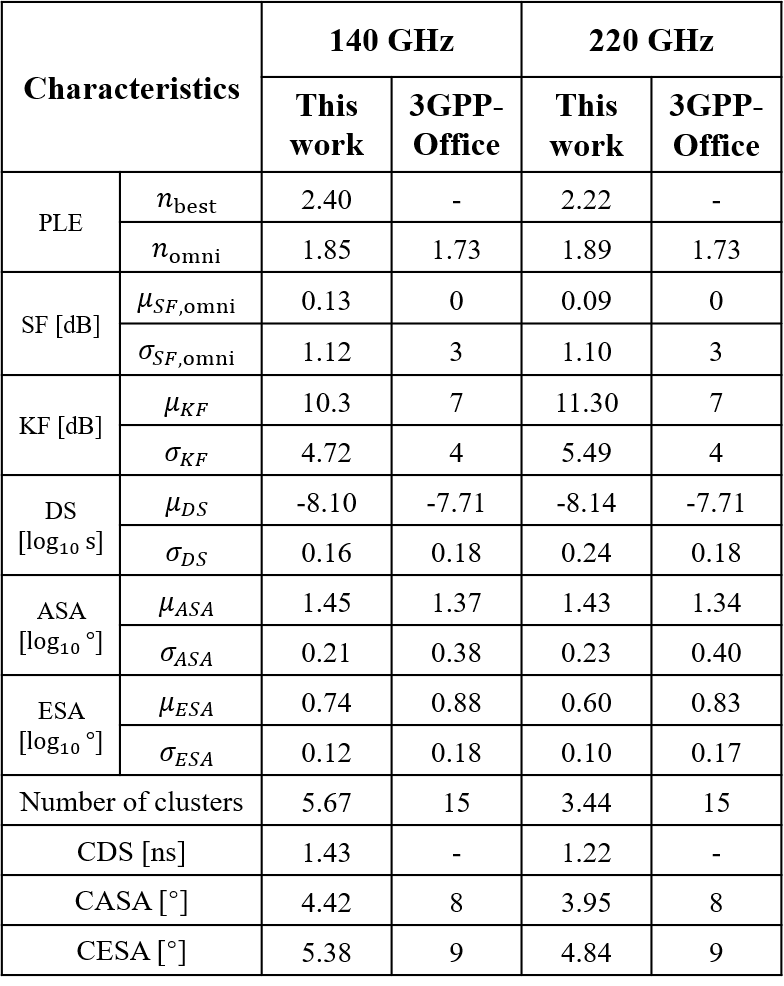}
    \label{tab:char}
    \vspace{-0.5cm}
\end{table}
\par The path loss is evaluated in terms of the best direction path loss and the omnidirectional path loss. The best direction path loss stands for the path loss on the steering direction of Rx with the strongest received power, while the omnidirectional path loss is related to the summation of received power from all MPCs, respectively as
\begin{align}
    \text{PL}_{\text{best}} ~[\text{dB}]&=-10\log_{10}(\max_{n_r}{\frac{1}{K}\sum_{k=1}^K\left|h_{n_r}[k]\right|^2})\\
    \text{PL}_{\text{omni}} ~[\text{dB}]&=-10\log_{10}(\sum_{l=1}^L{\left|\alpha_{l}\right|^2})
\end{align}
where $h_{n_r}$ is the CIR for the $(n_r)^\text{th}$ scanning direction of Rx. Additionally, $\alpha_l$ stands for the path gain of the $l^\text{th}$ MPC.
\par Moreover, the relationship between path loss and Tx-Rx distance can be modeled by using a close-in (CI) free space reference distance model, expressed as
\begin{equation}
\begin{split}
    \text{PL}^\text{CI}_{a} [\text{dB}]&=10n_{a}\log_{10}\frac{d}{d_0}+\text{FSPL}(d_0)+\chi_{a}
\end{split}
\end{equation}
where $a$ could either be “best” or “omni”. Besides, $n_{a}$ stands for the path loss exponent (PLE) of the CI model. Moreover, $d$ is the Euclidean distance between Tx and Rx. $d_0$ is the reference distance, set as \SI{1}{m} in this work. Moreover, $\chi_a$ is the Gaussian distributed shadow fading term. 
\par The PLEs and distribution parameters of shadow fading are shown in Table~\ref{tab:char}, based on which we can elaborate several observations as follows. First, the PLE of best direction path loss is larger than that of the omnidirectional path loss, which is reasonable as the best direction path loss only involves effects of partial MPCs while the omnidirectional path loss includes power of all MPCs. Second, the PLE of best direction path loss is slight larger than 2, which is the PLE of FSPL, attributed to misalignment of Rx antennas. In contrast, the PLE of omnidirectional path loss is slightly smaller than 2, due to the influences of multipath effects. Third, weak shadow fading effects are observed in the laboratory, with standard deviation around \SI{1}{dB}. Fourth, comparing the two frequency bands, the path loss and shadow fading model parameters are very close.
\subsection{K-factor}
\par The K-factor evaluates how dominant the strongest cluster is, which is calculated as the power ratio between the strongest cluster and remaining clusters. The measured K-factor values are fitted with log-normal distribution, where the distribution parameters are shown in Table~\ref{tab:char}. First, the mean K-factor values are \SI{10.3}{dB} and \SI{11.3}{dB} at \SI{140}{GHz} and \SI{220}{GHz}, respectively, indicating that the THz channels are dominant by the LoS cluster. Second, comparing the two frequency bands, the K-factor values slightly grow as the frequency increases, which is consistent with the observation in Sec.~\ref{sec:prop} that fewer clusters are received at \SI{220}{GHz} compared to \SI{140}{GHz}.
\subsection{Delay and Angular Spreads}
\par The power of MPCS disperses in both temporal and spatial domains, which can be characterized by delay and angular spreads. The delay spreads (DS), azimuth spreads of arrival (ASA) and the elevation spreads of arrival (ESA) are calculated and fitted with log-normal distributions, as shown in Table~\ref{tab:char}. Several observations are made as follows. First, the average delay spread values are \SI{7.94}{ns} and \SI{7.24}{ns} at \SI{140}{GHz} and \SI{220}{GHz}, respectively, which indicates that the temporal power dispersion in the laboratory is relatively weak. The slightly smaller delay spread value at \SI{220}{GHz} compared to that at \SI{140}{GHz} stems from less number of significant clusters. Second, the average angular spreads are $28.18^\circ$ and $5.50^\circ$ for ASA and ESA at \SI{140}{GHz}, respectively, while those values at \SI{220}{GHz} are $26.92^\circ$ and $3.98^\circ$. Similar to delay spread, the angular spreads slightly decrease as the frequency increase.
\subsection{Cluster Parameters}
\par The cluster parameters are calculated, including the number of clusters, cluster delay spread (CDS), cluster azimuth spread of arrival (CASA) and cluster elevation spread of arrival (CESA), whose average values are summarized in Table~\ref{tab:char}. The average number of clusters is 5.67 and 3.44 at \SI{140}{GHz} and \SI{220}{GHz}, respectively, indicating the sparsity of THz channels. Moreover, the mean CDS, CASA, and CESA are \SI{1.43}{ns}, $4.42^\circ$, and $5.38^\circ$ at \SI{140}{GHz}, respectively, while those values are \SI{1.22}{ns}, $3.95^\circ$ and $4.84^\circ$ at \SI{220}{GHz}, respectively. The intra-cluster delay and angular spreads also slightly reduce as the frequency increases.
\subsection{Comparison with Existing Channel Model}
\par The channel characteristics measured in this work are compared with the reference values from the existing channel model standardization file 3GPP TR 38.901~\cite{3gpp.38.901}, as shown in Table.~\ref{tab:char}. Note that as the frequency bands measured in this work are beyond the applicability of the 3GPP model, the parameters extrapolated from 3GPP model only shows what one may expect to see in the THz band. Since there is no laboratory scenario defined in 3GPP model, the parameters in the office scenario is selected here for comparison. Several take-away lessons are drawn as follows. First, the PLEs are slightly underestimated in the existing channel model, while the standard deviation of the shadow fading term is overestimated. Second, the measured K-factor values are larger than those predicated with 3GPP model, indicating that the THz channels are more dominant by the LoS cluster than one may expect by using 3GPP model. Third, the delay spread values are much smaller than that calculated using 3GPP model, revealing the weak multipath effects in the THz band. Fourth, the angular spread values from 3GPP model are close to those measured in this work, with the ASA slightly underestimated and ESA slightly overestimated. Last but not least, a significant difference between the cluster parameters measured in this work and those in 3GPP model is observed. The measured number of clusters appear smaller, for which the strong sparsity should be considered when conducting channel modeling in the THz band. Moreover, the intra-cluster spreads are also measured to be smaller. To summarize, compared to the predication from the existing 3GPP model, the multipath effects in realistic THz channels are weaker, for which strong sparsity and dominance of the LoS cluster emerges.
\section{Conclusion}
\label{sec:conclude}
In this paper, we conducted measurement campaigns in a laboratory at \SI{140}{GHz} and \SI{220}{GHz} using a correlation-based channel sounder. In the data post-processing procedures, the time drift of the clocks is corrected using a linear interpolation/extrapolation method. Based on the measured results, to examine how the objects in the laboratory affect the propagation of THz waves, we elaborated a thorough propagation by matching the once-scattering clusters to their propagation path, based on which the scattering losses are calculated. Furthermore, we calculated and analyzed the channel characteristics, including the path loss, shadow fading, K-factor, delay and angular spreads, as well as cluster parameters. Specifically, key observations are summarized as follows.
\begin{itemize}
    \item The main objects in the laboratory affecting the THz wave propagation are mostly made of metal. The multipath richness decreases as the frequency increases.
    \item The scattering losses from metal objects ranges from \SI{2}{dB} to \SI{28}{dB}, while the scattering losses from concrete walls and pillars are mostly within \SIrange{10}{25}{dB}.
    \item The measured THz channels appear sparse with weak multipath effects, i.e., very few clusters (less than 6) are observed with large K-factor values ($>$ \SI{10}{dB}).
\end{itemize}

\bibliographystyle{IEEEtran}
\bibliography{IEEEabrv,main}

\end{document}